\documentclass[global,final]{svjour}
\usepackage{epsfig}
\usepackage{amssymb,amsbsy,amsmath}
\newcommand{\fmref}[1]{(\protect\ref{#1})}

\def\ket#1{\, | \, {#1} \, \rangle}
\newcommand{\braket}[2]{\langle \, {#1} \, | \, {#2} \, \rangle}

\newcommand{\bi}[1]{\mathbf{#1}}

\journalname{Eur. Phys. J. B}
\begin{document}
\title{Improved upper and lower energy bounds for
  antiferromagnetic Heisenberg spin systems}
\titlerunning{Improved energy bounds}
\author{Klaus B\"arwinkel,  Heinz-J\"urgen Schmidt  \and J\"urgen Schnack
}
\offprints{J\"urgen Schnack}          
\institute{Universit\"at Osnabr\"uck, Fachbereich Physik,
Barbarastr. 7, D-49069 Osnabr\"uck, Germany}
\date{Received: date / Revised version: date}
%
\maketitle
\begin{abstract}
Large spin systems as given by magnetic macromolecules or
two-dimensional spin arrays rule out an exact diagonalization of
the Hamiltonian. Nevertheless, it is possible to derive upper
and lower bounds of the minimal energies, i.e. the smallest
energies for a given total spin $S$.

The energy bounds are derived under additional assumptions
on the topology of the coupling between the spins. The upper bound  follows
from ``n-cyclicity", which roughly means that the graph of interactions
can be wrapped round a ring with $n$ vertices. The lower bound
improves earlier results and
follows from ``n-homogeneity", i.~e.~from the assumption that the set
of spins can be decomposed into $n$ subsets where the
interactions inside and between spins of different subsets
fulfill certain homogeneity conditions. Many Heisenberg spin systems comply
with both concepts such that both bounds are available.

By investigating small systems which can be numerically diagonalized
we find that the upper bounds are considerably closer to the true
minimal energies than the lower ones.

\end{abstract}

\section{Introduction}
\label{s-1}

Rigorous results on spin systems like the Marshall-Peierls sign
rule \cite{Mar:PRS55} and the famous theorems of Lieb, Schultz,
and Mattis \cite{LSM:AP61,LiM:JMP62} have sharpened our
understanding of magnetic phenomena. In addition such results
can serve as a basis or source of inspiration for the
development of approximate models.  For example, the
inequalities of Lieb and Berezin \cite{Lie:CMP73,Ber:CMP75}
relating spectral properties of quantum systems to those of
their classical counterparts provide a foundation for classical
or semi-classical treatments of spin systems.

In this article we will extend the body of rigorous results on
Heisenberg spin systems by generalizing the notion of
``bi-partiteness", which is fundamental for the findings of
Marshall, Lieb, Schultz, and Mattis
\cite{Mar:PRS55,LSM:AP61,LiM:JMP62}.  We will introduce two new
concepts which rest on the topological properties of the
interaction matrix connecting the spins of the systems.

The first concept, n-cyclicity, uses the property of many spin
systems that their ``net" of interactions can be wrapped round an
$n$-cycle. The triangular lattice may serve as an example, it
can be mapped onto a triangle in an oriented manner as if one
would wrap it round the triangle. In such cases an upper bound
of the minimal energies $E_{{\text{min}}}$ in each subspace
${\cal H}(M)$ of total magnetic quantum number $M$ can be
derived for Heisenberg models and XXZ models in general.

The second concept which leads to lower bounds rests on
n-homogeneity, i.~e. on the fact that the set of spins can be
decomposed into $n$ subsets of equal size where the interactions
inside and between spins of different subsets fulfill certain
conditions.

Fortunately, many Heisenberg spin systems comply with both
concepts such that both bounds are available. For all cases
which were investigated it turns out that the upper bounds are
rather close to the true minimal energies, whereas the lower
bounds are not. Therefore, especially the upper bound can serve
are a benchmark or guideline for approximate methods like DMRG
or variational methods in order to rate the achieved quality.

The resulting bounds improve earlier findings of
Ref.~\cite{Lie:CMP73,Ber:CMP75,SSL:EPL01} especially for
frustrated spin systems.

The article is organized as follows. In section~\ref{s-2} upper
bounds and the concept of n-cyclicity will be discussed, in
section~\ref{s-3} lower bounds and n-homogeneity will be
introduced. Both sections start with subsections explaining the
idea followed by more mathematical subsections presenting the
mathematical tools. At the end of each section the resulting
bounds are given. Examples are provided in section~\ref{s-4}. A
more technical calculation is carried out in the appendix.

\section{Upper bounds}
\label{s-2}

\subsection{Idea}
\label{s-2-1}

It is obvious from the Ritz variational principle that an upper
bound for the minimal energy can be provided if an appropriate
trial state can be found for which the energy expectation value
is known analytically and rather close to the exact ground state
value. In the following we will construct such trial states for
subspaces ${\cal H}(M)$, i.~e. for total magnetic quantum number
$M$. Starting point is the magnon vacuum ($M=Ns$) which is
mapped by means of suitable powers of the total ladder operator
into the subspace ${\cal H}(M)$. Since the total ladder operator
commutes with the Heisenberg Hamilton operator this does not
change the energy of that state. In a second step we assume a
certain topological property of the spin array namely that it
can be wrapped round an $n$-cycle and construct a generalized
``Bloch operator" which is a unitary operator that adds
appropriate phases to the components of the trial
state. Utilizing the known action of the Bloch operator onto the
Hamiltonian we can evaluate the energy expectation value
analytically which results in the expression for the upper
bound.

In order to motivate our definitions in the next subsection we
recall the definition of a bi-partite spin system in the case of
constant coupling: It is required that the spin sites can be
grouped into $+$sites and $-$sites such that only $+-$ pairs are
coupled but no $++$ or $--$ pairs.  We suggest the following
generalization: Assume that complex phase factors $e^{i \phi_j}$
can be attached to spin sites $j$ such that only constant phase
differences $|\phi_j - \phi_k|$ occur between adjacent (coupled)
spin sites. This is the requirement needed for the
above-mentioned construction of the Bloch operator. The
attachment of phase factors is no longer arbitrary if there are
``loops" in the coupling scheme of the spin system,
i.~e.~periodic sequences of adjacent spin sites. If only even
loops exist we may choose the phase differences to be $|\phi_j -
\phi_k|=\pi$ and the system is bi-partite. However, in the case
of odd loops it becomes necessary to ``wrap" the loop around the
complex unit circle and the resulting phase differences will be
integer fractions of $2\pi$. We will make this more precise in
the next subsection employing the language of graph theory.

\subsection{Definition of n-cyclicity}
\label{s-2-2}

In this section we consider systems with $N$ spin sites with
spin $s$ and constant anti-ferromagnetic coupling.  Thus the
complete information about the coupling scheme is encoded in
some (undirected) graph $\gamma=({\cal V}, \Gamma)$. The
vertices of $\gamma$ are the spin sites, ${\cal
V}=\{1,\ldots,N\}$, the set of edges of $\gamma$ consists of
those pairs of sites which are coupled and will be denoted by
$\Gamma$. We make the convention that $\langle i, j \rangle \in
\Gamma$ iff $\langle j, i \rangle \in \Gamma$ and
$\langle i , i \rangle \notin \Gamma$.  Hence the number
of members of the set $\Gamma$, denoted by $|\Gamma|$, equals
twice the number of bonds.  Further, we will consider {\it
orientations} on $\gamma$, denoted by $\gamma^+$, i.~e.~we split
$\Gamma$ into disjoint subsets $\Gamma = \Gamma^+ \cup
\Gamma^-$, such that $\langle i, j \rangle \in \Gamma^+$ iff
$\langle j, i \rangle \in \Gamma^-$. Then the Hamiltonian of
$XXZ$-type can be written in the form
\begin{eqnarray}\label{1}
H
&=&
\delta \sum_{\langle i, j \rangle \in \Gamma} s_i^{(3)} s_j^{(3)}
+
\sum_{\langle i, j \rangle \in \Gamma^+} s_i^{+} s_j^{-}
+
\sum_{\langle i, j \rangle \in \Gamma^-} s_i^{+} s_j^{-}
\\
&\equiv&  \label{1a}
\Delta + {\cal G} +   {\cal G}^\dagger,
\end{eqnarray}
where $\delta>0$ and $\vec{s}_i$ denote the usual spin
observables at site $i$ with components $s_i^{(\mu)},\;
\mu=1,2,3$, and $s_i^{\pm}\equiv s_i^{(1)}\pm i s_i^{(2)}$.
Of course, only the splitting (\ref{1a}) depends on the orientation,
not the Hamiltonian itself.

In order to define a suitable concept of n-cyclicity we consider
{\it graph homomorphisms}, i.~e.~maps between graphs, such that
vertices are mapped onto vertices and the corresponding edges
onto corresponding edges. Let ${\cal C}_n$ denote the cyclic
graph with $n$ vertices which will be identified with the $n$-th
roots of unity
\begin{equation}\label{2}
e^{i \alpha_\ell} \equiv \exp
\left(
\frac{2 \pi i \ell}{n}
\right)
\ ,\qquad
\ell=0, \ldots, n-1
\ .
\end{equation}
Further  ${\cal C}_n^+$ will denote the cyclic graph with anti-clockwise
orientation.

Any graph $\gamma$ (or the spin system itself) will be called
n-cyclic or having the cyclicity $n$ iff there exists a graph
homomorphism 
\begin{equation}\label{3}
h: \gamma   \longrightarrow  {\cal C}_n
\ .
\end{equation}
In this case ${\cal C}_n^+$ will induce an orientation on
$\gamma$ in an obvious sense.

\begin{figure}[ht!]
\begin{center}
\epsfig{file=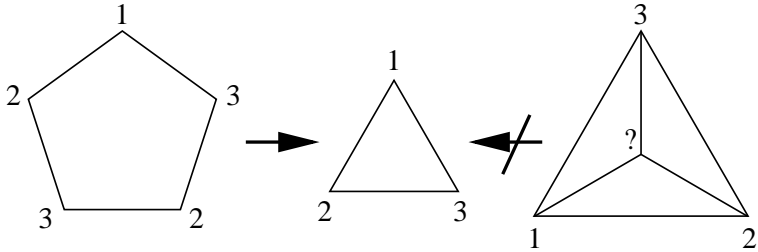,width=100mm}
\vspace*{1mm}
\caption[]{The pentagon is $5$-cyclic and also $3$-cyclic
  (l.h.s.) whereas the tetrahedron is not $3$-cyclic
  (r.h.s.), because if the four vertices of the tetrahedron are
  attached to the numbers $1,2,3$ one number must repeat and
  occurs at adjacent vertices, which does not happen in the
  $3$-cycle.}
\label{F-5}
\end{center}
\end{figure}
It is only in certain cases that different cyclicities $n$ and
$n^\prime$ mean an essential distinctness. This is because for
$n\ge 4$ any $n$-cyclic system is also $(n-2)$-cyclic since
three successive vertices and the corresponding edges can be
mapped in a forward-backward-forward way, compare the l.h.s. of
Fig.~\ref{F-5}, which shows a homomorphism of a pentagon onto a
triangle, as an example. Each $2m$-ring and hence any
$2m$-cyclic system is $n$-cyclic for any positive integers
$m,n$, since it is $2$-cyclic and ${\cal C}_2$ can be
homomorphically embedded into any $n$-cycle.

Hence it makes only sense to distinguish between even-cyclic
systems, which will be called $2$-cyclic, and $(2n+1)$-cyclic
system with maximal integer $n$.  If a spin system is $2$-cyclic
in our sense it will be bi-partite in the sense of
Refs.~\cite{LSM:AP61,LiM:JMP62}, where, however, the theory also
comprises cases with different coupling constants.

We consider some more examples which illustrate the definition
of cyclicity.  A triangular plane lattice with suitable periodic
boundary conditions is $3$-cyclic, a square lattice or cubic
lattice is $2$-cyclic.  The kagom\'e lattice is $3$-cyclic but
not $2$-cyclic. $3$-cyclicity is equivalent to
$3$-colorability. Hence the octahedron, the dodecahedron, the
cuboctahedron, and the icosidodecahedron are $3$-cyclic,
cf.~\cite{AxL:PRB01}, but the tetrahedron is not, see r.h.s. of
Figure~\ref{F-5}.

A natural basis for a matrix representation of $H$ is provided
by the product states $\ket{\mathbf m}= \ket{m_1 \ldots m_N}, \,
-s\le m_i \le s$ with
\begin{equation}\label{4}
s_i^{(3)} \ket{\mathbf m} = m_i \ket{\mathbf m}
\ ,
\end{equation}
and
\begin{equation}\label{5}
s_i^\pm \ket{\mathbf m} =\sqrt{s(s+1)-m_i(m_i\pm 1)}
\ket{m_1,\ldots, m_i\pm 1,\ldots,m_N}
\end{equation}
for all $i\in\{1, \ldots, N\}$.
The state $\ket{\Omega}\equiv\ket{s,s,\ldots,s}$ will be called the
``magnon vacuum". Further we define
\begin{equation}\label{6a}
a\equiv\sum_{i=1}^N a_i
=Ns-M
\ ,\quad
a_i = s - m_i\, \text{ for all } i\in\{1,\ldots,N\}
\ .
\end{equation}
We also define a mapping $\hat{h}$ of product states into
complex numbers which depends on the graph homomorphism
\fmref{3} by
\begin{equation}
\label{7b}
\hat{h}({\mathbf m})\equiv \prod_{i=1}^N h(i)^{a_i}
\ .
\end{equation}
Then it is easily shown that
if
$\langle  {\mathbf m} | {\cal G}|{\mathbf m}^\prime\rangle \neq 0$
then
$\hat{h}({\mathbf m}) = e^{2\pi i /n}\hat{h}({\mathbf
  m}^\prime)$.

For any $\ell=0,\ldots, N-1$ we define a unitary ``Bloch
operator" (generalizing the corresponding definition
for spin rings in Ref.~\cite{AfL:LMP86})
\begin{equation}\label{8a}
U_\ell: {\cal H}(M) \longrightarrow   {\cal H}(M)
\end{equation}
by
\begin{equation}\label{8b}
U_ \ell |\mathbf{m}\rangle = \hat{h} ({\mathbf m})^\ell
|\mathbf{m}\rangle
\end{equation}
and linear extension.  Recall that $\alpha_\ell = 2 \pi \ell/n$. Then
the following relations hold:
\begin{eqnarray} \label{9}
U_\ell^\dagger {\cal G} U_\ell &=& e^{-i\alpha_\ell}{\cal G}\\  \label{9a}
U_\ell^\dagger H U_\ell &=&
\Delta + \cos\alpha_\ell ({\cal G}+{\cal G}^\dagger)
-i \sin\alpha_\ell ({\cal G}-{\cal G}^\dagger)
\ .
\end{eqnarray}
If $E_{{\text{min}}}(M)$ denotes the minimal energy eigenvalue
within the sector ${\cal H}(M)$ and $\ket{\varphi}\in {\cal
H}(M)$ is an arbitrary normalized state we have the obvious
upper bound
\begin{equation}\label{10}
 E_{{\text{min}}}(M)   \le   \langle \varphi | H | \varphi
 \rangle
\ .
\end{equation}
The problem is to find a state $\ket{\varphi}$ such that
$\langle \varphi | H | \varphi \rangle$ can be explicitly
calculated and represents a close bound. To this end we map the
magnon vacuum $\ket{\Omega}$ by $(S^{-})^{a}$ into ${\cal
H}(M)$, which remains an eigenstate of $H$ with the largest
eigenvalue in the Heisenberg case $\delta=1$, and change its
phases according to the Bloch operator. More precisely, let
\begin{equation}\label{11}
|\Omega_M\rangle \equiv C_M (S^{-})^{a}   |\Omega\rangle
\ ,
\end{equation}
where $C_M$ is the positive normalization factor, compare
\fmref{13g}, ensuring 
\linebreak
$\braket{\Omega_M}{\Omega_M}=1$ and
define
\begin{equation}\label{12}
\ket{\varphi}\equiv \ket{\varphi_{M,\ell}} \equiv U_\ell
  |\Omega_M\rangle
\ ,
\end{equation}
Then we obtain
\begin{equation}\label{13}
\langle \Omega_M |\Delta |\Omega_M\rangle
=
\frac{\delta |\Gamma|}{N}
\left\{
Ns^2
-
\frac{2 s a ( 2 N s - a)}{2 N s - 1}
\right\}
\ .
\end{equation}
As it must be, this result has the obvious value $\delta
|\Gamma| s^2$ for $a=0$ and remains unchanged under $a
\leftrightarrow 2 N s - a$. The proof of Eq.~\fmref{13} is given
in appendix~\ref{a-1}.

Now consider
\begin{eqnarray}\label{14}
\langle \varphi |H|\varphi\rangle
&=&
\langle \Omega_M |U_\ell^{\dagger} H U_\ell|\Omega_M\rangle
\\   \label{14a}
&=&
\langle \Omega_M |\Delta+\cos\alpha_\ell
({\cal G}+  {\cal G}^{\dagger})
|\Omega_M\rangle \\   \label{14b}
&=&
\cos\alpha_\ell
\langle \Omega_M |\frac{\Delta}{\delta}+
({\cal G}+  {\cal G}^{\dagger})
|\Omega_M\rangle \\     \label{14c}
&&
+(1-\frac{\cos\alpha_\ell}{\delta})
\langle \Omega_M |\Delta | \Omega_M\rangle
\nonumber
\\   \label{14d}
&=&
|\Gamma| s^2  \cos\alpha_\ell
\\
&& +
(1-\frac{\cos\alpha_\ell}{\delta})
\frac{\delta |\Gamma|}{N}
\left(
N s^2 - \frac{2sa(2Ns-a)}{2Ns-1}
\right)
\nonumber
\ .
\end{eqnarray}
In line (\ref{14a}) we used (\ref{9a}) and the fact that
$\ket{\Omega_M}$ and ${\cal G}$ are real in the product basis of
the $|\mathbf{m}\rangle$ whence $\langle \Omega_M | ({\cal G} -
{\cal G}^{\dagger}) |\Omega_M\rangle=0$. Eq.~(\ref{14d}) follows
by Eq.~\fmref{13} and the observation that
$\frac{\Delta}{\delta}+ ({\cal G}+ {\cal G}^{\dagger})$ is a
Heisenberg Hamiltonian which has the eigenstate
$\ket{\Omega_M}$ with eigenvalue $|\Gamma | s^2 $.

For spin rings and $a=1$, $\ket{\varphi}$ is nothing else but
the relative ground state. Generally for spin rings,
$\ket{\varphi}$ has the same shift quantum number as the
relative ground state \cite{BHS:PRB03}.

\subsection{Analytical expression for the upper bound}
\label{s-2-3}

The best bound for $E_{{\text{min}}}(M)$
is obtained if $\cos\alpha_\ell$ in \fmref{14d} is as low as possible,
i.~e.~$\ell=\frac{n}{2}$ and $\cos\alpha_\ell=-1$ for even $n$ and
$\ell=\frac{n\pm 1}{2}$ for odd $n$.
Therefore the upper bound is given by
\begin{equation}\label{15}
E_{{\text{min}}}(M) \le
c |\Gamma| s^2
+
(1-\frac{c}{\delta})
\frac{\delta |\Gamma|}{N}
\left(
N s^2 - \frac{2sa(2Ns-a)}{2Ns-1}
\right)
\ ,
\end{equation}
where $c=-1$ in the case of even $n$ and $c=-\cos\frac{\pi}{n}$
for odd $n$.  Let $\delta=1$ and $Ns$ be integer. Then the total
ground state lies in the sector $M=0$. In this case we obtain
\begin{equation}\label{16}
E_{{\text{min}}}(0) \le c |\Gamma| s^2 +\frac{(-1+c)|\Gamma|
  s^2}{2Ns-1}
\ ,
\end{equation}
which improves the upper Berezin-Lieb bound
$E_{{\text{min}}}^{\text{classical}} s^2$, see
\cite{Lie:CMP73,Ber:CMP75}, if $c |\Gamma| =
E_{{\text{min}}}^{\text{classical}}$.

\section{Lower bounds}
\label{s-3}

\subsection{Idea}
\label{s-3-1}

For the lower bound to be derived in the following section the
interaction matrix ${\mathbb J}\equiv(J_{\mu\nu})$ describing
the coupling between spins at sites $\mu$ and $\nu$ must have
certain homogeneity properties. The matrix must be symmetric and
must have constant row sum. This alone is sufficient to derive
some lower bounds \cite{SSL:EPL01}, which can be improved using
the topological structure of the interactions, as will be shown
in the following.

The derivation works by constructing another ``averaged"
Hamiltonian having an analytically diagonalizable interaction
matrix $\widetilde{\mathbb J}$, which nevertheless has only
eigenvalues already present for the original interaction matrix
${\mathbb J}$. Since, by construction, ${\mathbb
J}\ge\widetilde{\mathbb J}$, this relation also holds for the
related Hamiltonians, and we arrive at a lower bound. Extending
this idea the obtained lower bounds could be improved for
particular systems. However, in this article we will confine
ourselves to deriving bounds for general classes of systems.

Using the topological structure of the interactions will further
enable us to determine the degeneracy of some eigenvalues of
${\mathbb J}$ and therefore improve the calculations of
Ref.~\cite{SSL:EPL01} where this information was not exploited.

\subsection{Definition of n-homogeneity}
\label{s-3-2}

The Hamiltonian used in this section is different from that of
the previous section and assumed to be of the form
\begin{equation}\label{2.1}
H=\sum_{\mu\nu}J_{\mu\nu}\vec{s}_\mu\cdot\vec{s}_\nu
\ .
\end{equation}
The matrix $\mathbb{J}$ of coupling constants $J_{\mu\nu}$ is
assumed to be symmetric and having constant row sums $j$. The
latter property can be viewed as a kind of gauge condition, since adding
a diagonal matrix with vanishing trace to  $\mathbb{J}$ does not
change the Hamiltonian (\ref{2.1}), see Ref.~\cite{SL:JPA03}.

Being symmetrical, $\mathbb{J}$ has a complete set of (ordered)
eigenvalues $j_1, \ldots, j_N$.  One of them is the row sum $j$
with $\bi{1}\equiv\frac{1}{\sqrt{N}}(1,1,\ldots,1)$ as the
corresponding eigenvector.  Let ${\mathbb J}^\prime$ denote the
matrix ${\mathbb J}$ restricted to the subspace orthogonal to
$\bi{1}$, and $j_{{\text{min}}}$ the smallest eigenvalue of
${\mathbb J}^\prime$.  $j_{{\text{min}}}$ may be $m$-fold
degenerate. Further, we will denote the $\alpha$-th normalized
eigenvector of ${\mathbb J}$ by
$(c_{1\alpha},\ldots,c_{N\alpha})$, i.~e.~
\begin{equation}\label{2.2}
\sum_{\nu}J_{\mu\nu} c_{\nu\alpha} = j_\alpha c_{\mu\alpha}
\ ,\
\sum_\mu \overline{c_{\mu\alpha}} c_{\mu\beta}=
\delta_{\alpha\beta}
\ ,\
\quad \alpha,\beta,\mu=1, \ldots, N
\ ,
\end{equation}
where we also allow for the possibility to choose complex
eigenvectors.  Sums over $\alpha=1,\ldots,N$ excluding
$\alpha_j$ will be denoted by $\sum'$, where $\alpha_j$ denotes
the index (within the ordered set of all eigenvalues) of the
eigenvalue $j$ belonging to the eigenvector $\bi{1}$.

For later use we will consider a transformation of the spin
observables analogous to the transformation onto the eigenbasis
of ${\mathbb J}$ and define
\begin{eqnarray}
{\vec{T}}_\alpha\equiv \sum_\mu \overline{c_{\mu\alpha}}
{\vec{s}}_\mu,
\text{ and  }
{Q}_\alpha\equiv{\vec{T}}_\alpha^\dagger\cdot {\vec{T}}_\alpha
\ , \quad \alpha=1, \ldots, N
\ .
\end{eqnarray}
The inverse transformation then yields
\begin{equation}\label{}
{\vec{s}}_\mu= \sum_\alpha c_{\mu\alpha}
{\vec{T}}_\alpha,\quad \mu=1, \ldots, N
\ .
\end{equation}
In particular, ${\vec{T}}_{\alpha_j}={\vec{S}}/\sqrt{N}$.
It then follows directly from the definitions that
\begin{equation}
\label{2.3}
Ns(s+1)
=
\sum_{\mu} ({\vec{s}}_\mu)^2
=
\sum_\alpha {Q}_\alpha
=
\frac{1}{N}{\vec{S}}^2
+ \sum_\alpha{}^\prime\; {Q}_\alpha
\ ,
\end{equation}
\begin{equation}
\label{2.4}
{H}
=
\sum_{\mu\nu\alpha\beta} J_{\mu\nu} \overline{c_{\mu\alpha}}c_{\nu\beta}
{\vec{T}}_\alpha^\dagger\cdot {\vec{T}}_\beta
=   \sum_\beta j_\beta {Q}_\beta
=  \frac{j}{N} {\vec{S}}^2 +  \sum_\beta{}^\prime\; j_\beta
{Q}_\beta
\ .
\end{equation}
For a later use we also need a relation between Hamiltonians
with different coupling matrices. Therefore, let $H$ and
$\widetilde{H}$ be two Hamiltonians of the form (\ref{2.1}) with
coupling matrices ${\mathbb J}$ and $\widetilde{\mathbb J}$,
such that ${\mathbb J}\ge \widetilde{\mathbb J}$. Then $H \ge
\widetilde{H}$.
As usual the ordering ``$\ge$" of operators is defined by the
corresponding inequality for arbitrary expectation values.
Since $H$ depends linearly on ${\mathbb J}$ it suffices to show
that ${\mathbb J}\ge 0$ implies $H\ge 0$. But this is obvious in
view of \fmref{2.4}: $H=\sum_\beta j_\beta {Q}_\beta$ with
$j_\beta\ge 0$ and $Q_\beta\ge 0$.

Next we turn to the suitable definition of n-homogeneity.  Let
the set of spin sites $\{1, \ldots, N\}$ be divided into $n$
disjoint subsets of equal size $m$, $\{1, \ldots, N\}=
\bigcup_{\nu=1}^{n}{\cal A}_\nu$, such that the coupling
constants within each ${\cal A}_\nu$ are $\le 0$, but $\ge 0$
between ${\cal A}_\nu$ and ${\cal A}_\mu$ for $\nu\neq\mu$.
Moreover, the partial row sums are assumed to be constant:
\begin{equation}\label{2.5}
\sum_{b\in{\cal A}_\mu}J_{ab}=\left\{
\begin{array}{lcl}
j^{\text{in}} & \text{if} & a\in {\cal A}_\mu \\
j^{\text{ex}} & \text{if} & a\not\in {\cal A}_\mu
\end{array}
\right.
\ .
\end{equation}
A spin system satisfying the assumptions of this section will be
called n-homogeneous, see \cite{Mar:PRS55,LSM:AP61,LiM:JMP62}.
Note that this notion is incommensurable to n-cyclicity defined
in the previous section. However, certain rings, the triangular
lattice, the kagom\'e lattice, and the icosidodecahedron satisfy
both definitions. A necessary condition for nearest neighbor
Heisenberg systems to be n-homogeneous is that the number of
nearest neighbors, which is assumed to be constant, is divisible
by $(n-1)$. Actually, spin rings of even $N$ are
$2$-homogeneous, rings of odd $N$ are $3$-homogeneous if $N$ is
divisible by 3. n-homogeneous Heisenberg rings do not exist for $n>3$
because they do not fulfill the homogeneity condition
\fmref{2.5}.

We recall that $\bi{1}=\frac{1}{\sqrt{N}}(1,1,\ldots,1)$ is an
eigenvector of ${\mathbb J}$ with eigenvalue $j$.  Due to
n-homogeneity there are, after a suitable permutation of the
spin sites, further eigenvectors of the form
\begin{equation}\label{2.6}
u^{(k)}=(m:1,m:\rho^{k},m:\rho^{2k},\ldots,m:\rho^{(n-1)k}),
\; k=1, \ldots, n-1
\ ,
\end{equation}
where $(m:x,\ldots)$ denotes the $m$-fold repetition of the
entry $x$, and $\rho\equiv e^{2\pi i/n}$. The corresponding
eigenvalues are
$j_k=j^{\text{in}}+j^{\text{ex}}\sum_{p=1}^{n-1}\rho^{pk}
=j^{\text{in}}-j^{\text{ex}}$, hence they coalesce into one
$(n-1)$-fold degenerate eigenvalue. By applying the theorem of
Ger\v{s}gorin (c.f.~\cite{Lan69}, 7.2) this eigenvalue is shown
to be the smallest one $j_{{\text{min}}}$.

Next we construct a coupling matrix $\widetilde{\mathbb J}$ with
the same eigenspaces as ${\mathbb J}$ but only three different
eigenvalues.  It has the block structure
\begin{equation}
\widetilde{\mathbb J}=
\left(
\begin{array}{cccc}
A & C & C & \ldots\\
C & A & C & \ldots\\
C & C & A & \ldots\\
\vdots & \vdots &\vdots &\vdots
\end{array}
\right)
\ ,
\end{equation}
where $A$ and $C$ are $m\times m$-matrices of the form
\begin{equation}
A=
\left(
\begin{array}{cccc}
\beta & -\alpha & -\alpha & \ldots\\
-\alpha & \beta &-\alpha& \ldots\\
-\alpha & -\alpha & \beta & \ldots\\
\vdots & \vdots &\vdots &\vdots
\end{array}
\right)
\ ,\quad
C=
\left(
\begin{array}{cccc}
\gamma & \gamma & \gamma & \ldots\\
\gamma & \gamma &\gamma& \ldots\\
\gamma & \gamma & \gamma & \ldots\\
\vdots & \vdots &\vdots &\vdots
\end{array}
\right)
\ .
\end{equation}
The three eigenvalues of $\widetilde{\mathbb J}$ are
\begin{eqnarray}\label{2.7}
\tilde{\jmath}
&=&
\beta - (m-1)\alpha +(N-m)\gamma,\\
\tilde{\jmath}_{{\text{min}}}
&=&
\beta - (m-1)\alpha -m \gamma,\\
\tilde{\jmath}_2
&=&
\alpha+\beta
\end{eqnarray}
with degeneracies $1, n-1$ and $N-n$, resp. By choosing
\begin{eqnarray}\label{2.8}
\alpha
&=&
\frac{1}{N}(n j_2-j-(n-1)j_{\text{min}}),\\
\beta
&=&
\frac{1}{N}((N-n)j_2+j+(n-1)j_{\text{min}}),\\
\gamma
&=&
\frac{j-j_{\text{min}}}{N}
\ ,
\end{eqnarray}
one obtains
\begin{equation}\label{2.9}
\tilde{\jmath}=j,\;
\tilde{\jmath}_{\text{min}}=j_{\text{min}},\;
\tilde{\jmath}_2=j_2
\ .
\end{equation}
$j_2$ is the remaining smallest eigenvalue of ${\mathbb
J}{}^\prime$ after eliminating $(n-1)$-times $j_{\text{min}}$
from the set of eigenvalues. Thus it can happen that
$j_2=j_{\text{min}}$ if $j_{\text{min}}$ is more than
$(n-1)$-fold degenerate.

Let us write $\vec{S}_{\cal A} \equiv \sum_{a\in{\cal A}}  \vec{s}_a$
for any subset ${\cal A}\subset\{1,\ldots,N\}$. We conclude
\begin{eqnarray}\label{2.10}
H
&\ge&
\widetilde{H} =
-\alpha\left(\sum_\nu \vec{S}_{{\cal A}_\nu}^{2}\right)
+ (\alpha + \beta) N s(s+1) +
\gamma
\left(\vec{S}^{2}-\sum_\nu \vec{S}_{{\cal A}_\nu}^{2}\right) \\  \label{2.10a}
&\ge&
\gamma S(S+1) - (\alpha+\gamma)n \frac{N}{n}s(\frac{N}{n}s+1)
+(\alpha+\beta) N s(s+1)
\ .
\end{eqnarray}

\subsection{Analytical expression for the lower bound}
\label{s-3-3}

Hence we obtain for the lower bound
\begin{eqnarray}\label{2.10b}
E
&\ge&
\frac{j-j_{\text{min}}}{N}S(S+1)+N  j_{\text{min}} s(s+1)
+(N-n) (j_2-j_{\text{min}}) s
\ .
\end{eqnarray}
Since $j_2-j_{\text{min}} \ge 0$ the bound (\ref{2.10b})
is the better, the smaller $n$ is. This is in contrast to the
upper bound considered in the previous section, which is improved
for large odd $n$.

\section{Examples}
\label{s-4}

In the following examples we calculate the energy eigenvalues by
numerical methods as well as lower and upper bounds. All
examples are Heisenberg spin systems where the total spin $S$ is
a good quantum number.  It turns out that $S\mapsto
E_{\text{min}}(S)$ is always a monotonically increasing
function, hence we need not to distinguish between
$E_{\text{min}}(S)$ and $E_{\text{min}}(M)$.

In order to judge the quality of the bounds we provide the
deviation of the best upper and lower bound from the exact
ground state energy in relation to the energy difference between
antiferromagnetic and ferromagnetic ground state, i.~e.
\begin{eqnarray}\label{4.1}
\epsilon
&=&
\frac{|E_{\text{bound},0}-E_0|}{E(Ns)-E_0}
\ .
\end{eqnarray}

\begin{figure}[ht!]
\begin{center}
\epsfig{file=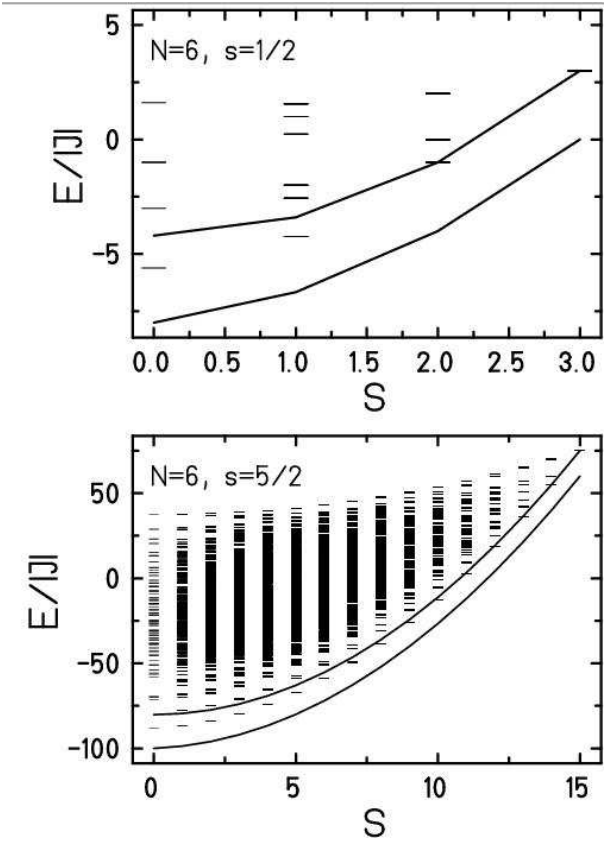,width=80mm}
\vspace*{1mm}
\caption[]{Upper and lower bounds of $E_{\text{min}}(S)$ for
  Heisenberg spin rings with $N=6$ and $s=1/2$ (top) as well
  as $s=5/2$ (bottom). The solid curves display the bounds for
  the minimal energies considering $2$-cyclicity ($s=1/2:
  \epsilon=0.16; s=5/2: \epsilon=0.05$)
  and $2$-homogeneity ($s=1/2: \epsilon=0.28; s=5/2:
  \epsilon=0.07$).}
\label{F-1}
\end{center}
\end{figure}
The first example we would like to consider is a Heisenberg spin
ring with $N=6$ and $s=1/2$ as well as $s=5/2$. Figure~\ref{F-1}
shows the numerically determined energy eigenvalues (dashes) as
a function of total spin $S$. The solid curves display the
bounds for the minimal energies considering $2$-cyclicity and
$2$-homogeneity.

\begin{figure}[ht!]
\begin{center}
\epsfig{file=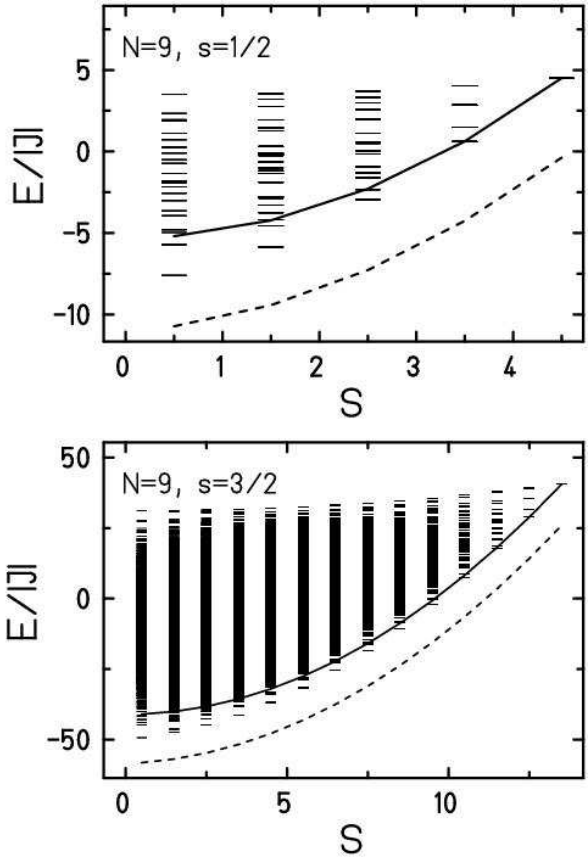,width=80mm}
\vspace*{1mm}
\caption[]{Upper and lower bounds of $E_{\text{min}}(S)$ for
  Heisenberg spin rings with $N=9$ and $s=1/2$ (top) as well
  as $s=3/2$ (bottom). The solid curves display the upper bounds
  for the minimal energies considering $9$-cyclicity ($s=1/2:
  \epsilon=0.20; s=3/2: \epsilon=0.09$), the
  dashed curves do the same for $3$-homogeneity ($s=1/2:
  \epsilon=0.26; s=3/2: \epsilon=0.10$).} 
\label{F-2}
\end{center}
\end{figure}
As a second example we take a frustrated Heisenberg ring with
$N=9$ and $s=1/2$ as well as $s=3/2$. The results are presented
in Fig.~\ref{F-2}. The solid curves display the upper bounds for
the minimal energies considering $9$-cyclicity, the dashed
curves do the same for $3$-homogeneity. Without using the
concept of $n$-homogeneity the lower bounds are much poorer for
frustrated systems \cite{SSL:EPL01}.

\begin{figure}[ht!]
\begin{center}
\epsfig{file=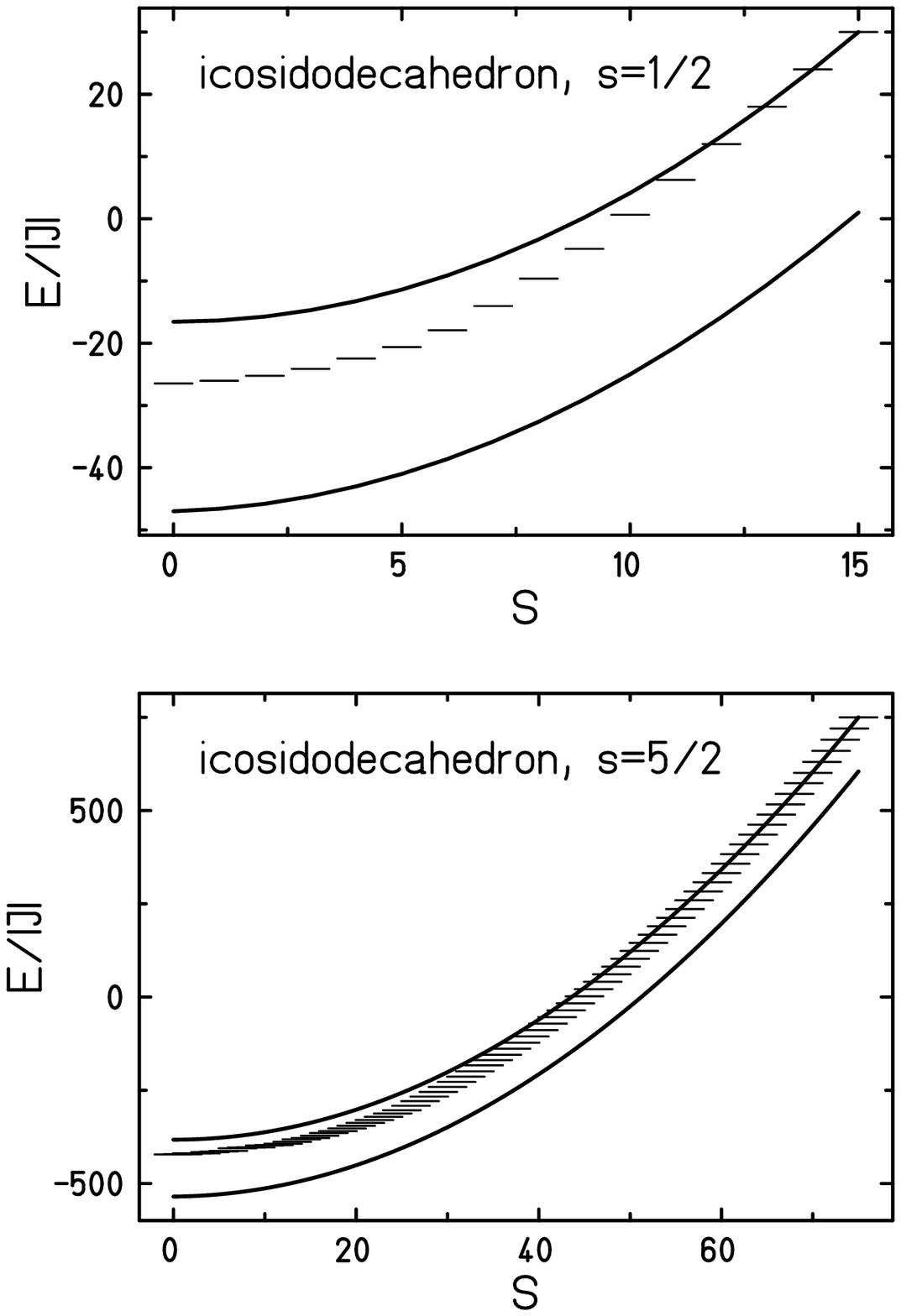,width=80mm}
\vspace*{1mm}
\caption[]{Upper and lower bounds of $E_{\text{min}}(S)$ for
  Heisenberg spin systems with icosidodecahedral structure,
  i.~e.~ $N=30$ and $s=1/2$ (top) as well as $s=5/2$
  (bottom). The solid curves display the bounds for the minimal
  energies considering $3$-cyclicity ($s=1/2: \epsilon=0.18;
  s=5/2: \epsilon=0.03$) and
  $3$-homogeneity ($s=1/2: \epsilon=0.36; s=5/2: \epsilon=0.10$).}
\label{F-3}
\end{center}
\end{figure}
Another example, an icosidodecahedral Heisenberg spin system, is
related to magnetic molecules, which can be synthesized in such
structures. One species is given by \{Mo$_{72}$Fe$_{30}$\}, a
molecule where 30 Fe$^{3+}$ paramagnetic ions ($s=5/2$) occupy
the sites of a perfect icosidodecahedron \cite{polytopes} and
interact via isotropic nearest-neighbor antiferromagnetic
Heisenberg exchange \cite{MSS:ACIE99}. Not much is known about
the spectrum of such giant structures since the Hilbert space
assumes a very large dimension of $6^{30}\approx 10^{23}$. So
far only DMRG calculations could approximate the minimal
energies \cite{MeS:PRB03}.

Figure~\ref{F-3} shows as dashes on the l.h.s. the minimal
energies for $s=1/2$ which are determined numerically by
J.~Richter with a L\'anczos method \cite{SSR:EPJB01,Ric:PC} and on
the r.h.s. the minimal DMRG energies \cite{MeS:PRB03}.  The
icosidodecahedral Heisenberg spin system is $3$-cyclic as well
as $3$-homogeneous. The corresponding bounds are displayed by
solid curves. Especially the upper bound for the case of $s=5/2$
is very close to the ``true" (DMRG) minimal energies and thus
could be used to justify approximations of the low-lying
spectrum as used in Ref.~\cite{SLM:EPL01}. The lower bounds are
worse than expected, but this behavior is explained by the
10-fold degeneracy of $j_{\text{min}}$, therefore
$j_2=j_{\text{min}}$, and the last term in \fmref{2.10b} yields
zero, unfortunately.

\begin{figure}[ht!]
\begin{center}
\epsfig{file=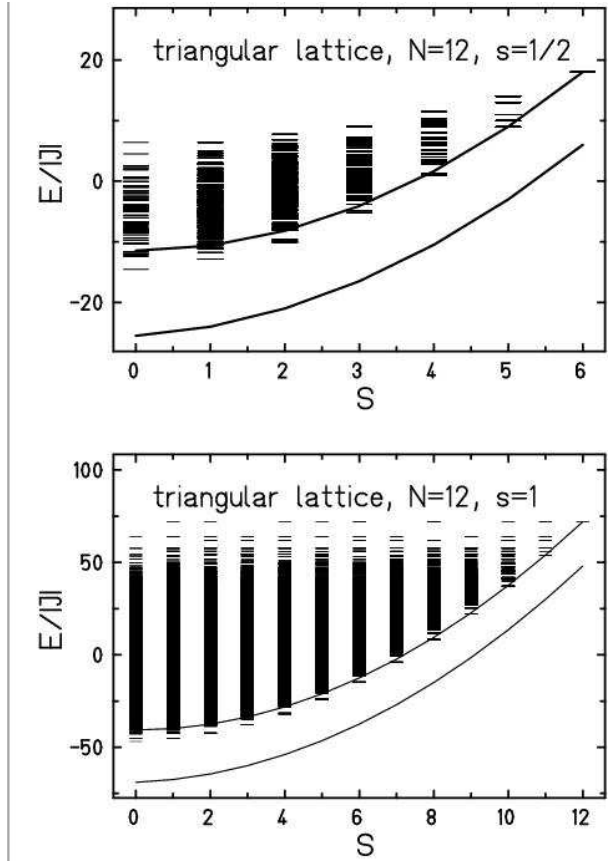,width=80mm}
\vspace*{1mm}
\caption[]{Upper and lower bounds of $E_{\text{min}}(S)$ for the
  triangular spin lattice with $N=12$ and $s=1/2$ (top) as
  well as $s=1$ (bottom). The solid curves display the bounds
  for the minimal energies considering $3$-cyclicity ($s=1/2:
  \epsilon=0.09; s=1: \epsilon=0.05$) and
  $3$-homogeneity ($s=1/2: \epsilon=0.34; s=1: \epsilon=0.19$).}
\label{F-4}
\end{center}
\end{figure}
The last example discusses the triangular spin lattice which is
one of the frustrated two-dimensional spin systems. The
triangular spin lattice is $3$-homogeneous and $3$-cyclic, if
the periodic boundary conditions are suitably chosen.
Figure~\ref{F-4} displays the energy levels for $N=12$ and
$s=1/2$ (l.h.s.) as well as $s=1$ (r.h.s.). The bounds of
$E_{\text{min}}(S)$ are given by solid curves. In both cases the
upper bound is very close to the exact minimal energies.

For the thermodynamic limit $N\rightarrow \infty$ of the
triangular lattice with $\delta=1$ we rewrite the bounds by
introducing a continuous spin variable $S_c=S/N$ running from
$0$ to $s$ and using $j=6$, $j_{\text{min}}=-3$ with twofold
degeneracy, and $\lim_{N\rightarrow\infty} j_2=-3$.  After
dividing by $N$ the resulting bounds are separated only by $3s$:
\begin{eqnarray}
\label{E-3}
9 S_c^2
-3 s^2
-3 s
\le
\lim_{N\rightarrow\infty}
\frac{E_{\text{min}}(S)}{N}
\le
9 S_c^2 - 3 s^2
\ .
\end{eqnarray}

\appendix
\section{Proof of Eq. \fmref{13}}
\label{a-1}

For easy readability we repeat Equation~\fmref{13}
\begin{equation}
\langle \Omega_M |\Delta |\Omega_M\rangle
=
\frac{\delta |\Gamma|}{N}
\left\{
Ns^2
-
\frac{2 s a ( 2 N s - a)}{2 N s - 1}
\right\}
\ .
\end{equation}

Since $\Omega_M$ is invariant w.~r.~t.~arbitrary permutations of
spin sites it suffices to choose $\Delta= s_1^{(3)} s_2^{(3)}$
and to multiply the result for
$\langle \Omega_M |\Delta |\Omega_M\rangle$
by $\delta |\Gamma|$. We note that
\begin{equation}\label{13a}
\left[\Delta,S^- \right]  = -( s_1^{(3)} s_2^{-}+ s_1^{-} s_2^{(3)}),
\end{equation}
and
\begin{equation}\label{13b}
\left[\left[\Delta,S^-\right],S^-\right]=2  s_1^{-} s_2^{-},
\end{equation}
but higher commutators vanish. Hence
\begin{equation}\label{13c}
\left[\Delta,(S^-)^a\right]
=
a (S^-)^{a-1}  \left[\Delta,S^- \right]  +
{a\choose 2} (S^-)^{a-2}
\left[\left[\Delta,S^-\right],S^-\right]
\ .
\end{equation}
Further we define $\lambda(a,k)$ by
\begin{equation}\label{13d}
(S^+)^a  (S^-)^a  (S^-)^k \ket{\Omega} = \lambda(a,k)  (S^-)^k \ket{\Omega}
\ .
\end{equation}
Using $ S^+ S^- = S^2 -S^{(3)}(S^{(3)}-1)$ one derives the
recursion relation
\begin{equation}\label{13e}
\lambda(a+1,k)=(2Ns-a-k)(a+k+1)\lambda(a,k)
\ .
\end{equation}
Together with $\lambda(0,k)=1$ it can be solved and yields
\begin{equation}\label{13f}
\lambda(a,k)=\frac{(2Ns-k)!}{(2Ns-a-k)!}\frac{(a+k)!}{k!}
\ .
\end{equation}
Obviously,
\begin{equation}\label{13g}
C_M^2=\lambda(a,0)^{-1}=\frac{(2Ns-a)!}{(2Ns)!\ a!}
\ ,
\end{equation}
hence
\begin{equation}\label{13h}
C_M^2 \lambda(a-1,1)=\frac{(2Ns-a)!}{(2Ns)!\ a!} \frac{(2Ns-1)!\ a!}{(2Ns-a)!\ 1!}
= \frac{1}{2Ns}
\ ,
\end{equation}
and
\begin{equation}\label{13i}
C_M^2 \lambda(a-2,2)=\frac{(2Ns-a)!}{(2Ns)!\ a!} \frac{(2Ns-2)!\ a!}{(2Ns-a)!\ 2!}
= \frac{1}{4Ns(2Ns-1)}
\ .
\end{equation}
Now we are prepared to calculate $\langle \Omega_M |\Delta |\Omega_M\rangle$:
\begin{eqnarray}\label{13j}
\langle \Omega_M |\Delta |\Omega_M\rangle
&=&
C_M^2 \langle \Omega_M |
(S^-)^a\Delta + \left[ \Delta,(S^-)^a\right]
|\Omega\rangle
\\
&=&
s^2 +
a C_M^2   \langle \Omega_M |
(S^-)^{a-1} \left[ \Delta,S^-\right]
|\Omega\rangle
\label{13k}
\\
&+&
{a\choose 2}
C_M^2   \langle \Omega_M |
(S^-)^{a-2} \left[\left[ \Delta,S^-\right] ,S^-\right]
|\Omega\rangle
\nonumber
\\
&=&
s^2 +
a C_M^2   \langle (S^+)^{a-1} (S^-)^{a-1}S^-\Omega |
 \left[ \Delta,S^-\right]
|\Omega\rangle
 \label{13l}
\\
&+&
{a\choose 2}
C_M^2   \langle (S^+)^{a-2} (S^-)^{a-2}(S^-)^{2}\Omega |
\left[\left[ \Delta,S^-\right] ,S^-\right]
|\Omega\rangle
\nonumber
\\
&=&
s^2 +
a C_M^2  \lambda(a-1,1) \langle S^-\Omega |
 \left[ \Delta,S^-\right]
|\Omega\rangle
\label{13m}
\\
&+&
{a\choose 2}
C_M^2   \lambda(a-2,2) \langle (S^-)^{2}\Omega |
\left[\left[ \Delta,S^-\right] ,S^-\right]
|\Omega\rangle
\nonumber
\\
&=&
\label{13n}
s^2 +
\frac{a}{2Ns}(-4 s^2)  +{a\choose 2} \frac{1}{4Ns(2Ns-1)} 16s^2  \\
&=&
\frac{1}{N}
\left\{
Ns^2
-
\frac{2 s a ( 2 N s - a)}{2 N s - 1}
\right\}
\ .
\end{eqnarray}
In line (\ref{13k}) we used (\ref{13c}). (\ref{13n}) is obtained
by means of ({\ref{13h}), (\ref{13i}) and the identities
\begin{eqnarray}\label{13o}
\langle S^-\Omega |
 \left[ \Delta,S^-\right]
|\Omega\rangle
&=&
\langle S^-\Omega | -( s_1^{(3)} s_2^{-}+ s_1^{-} s_2^{(3)})   |\Omega\rangle
\\
&=&
-2 \langle s_2^{-}\Omega |  s_1^{(3)} s_2^{-}  |\Omega\rangle
= -2 s (s(s+1)-s(s-1))
\nonumber
\\
&=&-4s^2\ ,
\nonumber
\end{eqnarray}
and
\begin{eqnarray}\label{13p}
\langle (S^-)^{2}\Omega |
\left[\left[ \Delta,S^-\right] ,S^-\right]
|\Omega\rangle
&=&
\langle (S^-)^{2}\Omega | 2 s_1^{-} s_2^{-} |\Omega\rangle
\\
&=&
4\langle s_1^{-} s_2^{-}\Omega | s_1^{-} s_2^{-} |\Omega\rangle
\nonumber
\\
&=&
4 (s(s+1)-s(s-1))^2
\nonumber
\\
&=&
16s^2\ .
\nonumber
\end{eqnarray}
This completes the proof.
\hfill  $\blacksquare$

\begin{acknowledgement}
We thank Johannes Richter for fruitful discussions.
\end{acknowledgement}


\end{document}